\documentclass[10pt] {IEEEtran} 
\usepackage{booktabs}
\usepackage{cite}
\usepackage[dvips]{graphicx}
\usepackage[cmex10]{amsmath}
\usepackage{amsfonts} 
\usepackage{array}
\usepackage{mdwmath}
\usepackage{mdwtab}
\usepackage[font=footnotesize, caption=false]{subfig}
\usepackage{fixltx2e}
\usepackage{soul}
\usepackage{verbatim}
\usepackage{url, epstopdf}

\usepackage{color}
\usepackage[ruled]{algorithm2e}

\makeatletter
\renewcommand*\env@matrix[1][c]{\hskip -\arraycolsep
  \let\@ifnextchar\new@ifnextchar
  \array{*\c@MaxMatrixCols #1}}
\makeatother

\begin{document}

\title{Distributed Widely Linear Complex Kalman Filtering}
\author{%
{Dahir H. Dini, Sithan Kanna and Danilo P. Mandic}
%
\thanks{Dahir H. Dini, Sithan Kanna and Danilo P. Mandic are with the Department of Electrical and Electronic Engineering, Imperial College London, Exhibition Road, London, SW7 2BT, UK. (e-mail: $\{$dahir.dini, ssk08, d.mandic$\}$@imperial.ac.uk).}
}
%
%
%
%
\maketitle
%
\begin{abstract}
We introduce cooperative sequential state space estimation in the domain of augmented complex statistics, whereby nodes in a network collaborate locally to estimate noncircular complex signals. For rigour, a distributed augmented (widely linear) complex Kalman filter (D-ACKF) suited to the generality of complex signals is introduced, allowing for unified treatment of both proper (rotation invariant) and improper (rotation dependent) signal distributions. Its duality with the bivariate real-valued distributed Kalman filter, along with several issues of implementation are also illuminated. The analysis and simulations show that unlike existing distributed Kalman filter solutions, the D-ACKF caters for both the improper data and the correlations between nodal observation noises, thus providing enhanced performance in real-world scenarios. 
\end{abstract}
%
%
\section{Introduction}
Distributed estimation and fusion has received significant attention in both military and civilian applications \cite{Stadter_Dist_Spacecraft_IEEEMag_2002, Mandic_Fusion_Book_2008, Cattivelli_Sayed_IEEETranAC_Dist_KF_2010, Olfati_Saber_Flocking_2006}, and the recent advances in sensor technology and wireless communications have highlighted the usefulness of distributed networks in this context \cite{Zhou_Dist_Architectures_IEEEMag_2003,Stadter_Dist_Spacecraft_IEEEMag_2002}. Such models rely on cooperation between the nodes (sensors) to provide more accurate and robust estimation compared to using independent uncooperative nodes, while approaching the performance of the more complex centralised systems. This is achieved through nodes equipped with learning capabilities that take local measurements (observations) and share information with their neighbours, thus enhancing robustness to link and node failures and facilitating scalability \cite{Olfati_Saber_Dist_KF,Carli_Dist_KF_IEEEComm_2008,Khan_Dist_KF_IEEETransSP_2008}. A number of robust and scalable diffusion strategies for network cooperation have been developed for distributed least-mean-square estimation \cite{Lopes_Sayed_TSP_Dist_LMS_2008, Yili_Dist_ACLMS_2011} and Kalman filtering  \cite{Olfati_Saber_Dist_KF, Cattivelli_Sayed_IEEETranAC_Dist_KF_2010}, however, these are linked to a very restrictive class of proper signals, and are also inadequate for correlated measurement noises, a common case in practice. 

Complex signals arise in a number of distributed real-world applications, such as in wireless communication systems \cite{Gao_Dist_Cooperative_IEEETranComm_2011, Mao_Wireless_Comm_IEEETranIFS_2007} and power systems \cite{Xia_WL_Freq_2011}. However, standard complex algorithms are generally suboptimal unless the underlying signals are proper (circular), that is, with rotation invariant probability distributions \cite{Picinbono97, Moreno_2008}. For a zero-mean proper complex signal, $\mathbf{x}$, the covariance matrix, $\mathbf{R}_{\mathbf{x}}=E\{\mathbf{x}\mathbf{x}^H\}$ suffices to represent its complete second-order statistics. For improper (noncircular) complex signals, another statistical moment function known as the pseudocovariance, $\mathbf{P}_{\mathbf{x}}=E\{\mathbf{x}\mathbf{x}^T\}$, which captures the information about the power difference and cross-correlation between the real and imaginary parts of the signal, is also required for a full second order statistical description \cite{Picinbono97}. 

To illuminate this further, consider the  minimum mean square error (MSE) estimator of a zero-mean real valued random vector $\mathbf{y}$ in terms of an observed zero-mean real vector $\mathbf{x}$, that is, $\hat{\mathbf{y}} = E\{\mathbf{y}|\mathbf{x}\}$. For jointly normal $\mathbf{y}$ and $\mathbf{x}$, the optimal linear estimator is 
\begin{equation}
	\hat{\mathbf{y}} = \mathbf{A}\mathbf{x} 
	\label{Eq:Linear_Estimator}
\end{equation}
where $\mathbf{A} =\mathbf{R}_{\mathbf{yx}}\mathbf{R}_{\mathbf{x}}^{-1}$ is a coefficient matrix, and $\mathbf{R}_{\mathbf{yx}} = E\{\mathbf{y}\mathbf{x}^H\}$. Standard, `strictly linear' estimation in ${\mathbb C}$ assumes the same model but with complex valued $\mathbf{y}, \mathbf{x}$, and $\mathbf{A}$. However, when $\mathbf{y}$ and $\mathbf{x}$ are jointly improper $\mathbf{P}_{\mathbf{yx}} = E\{\mathbf{y}\mathbf{x}^T\} \neq \mathbf{0}$, and $\mathbf{x}$ is improper so that $\mathbf{P}_{\mathbf{x}} \neq \mathbf{0}$, then the optimal estimator becomes widely linear\footnote{The `widely linear' model is associated with the signal generating system, whereas ``augmented statistics'' describe statistical properties of measured signals. Both the terms `widely linear' and `augmented' are used to name the resulting algorithms - in our work we mostly use the term `augmented'.}, that is \cite{Picinbono97}
\begin{equation}
	\hat{\mathbf{y}} = \mathbf{B}\mathbf{x} + \mathbf{C}\mathbf{x}^*  
									 = \mathbf{W} \mathbf{x}^{a}
	\label{Eq:WL_model}
\end{equation}
where $\mathbf{B} = \mathbf{R}_{\mathbf{yx}}\mathbf{D} + \mathbf{P}_{\mathbf{yx}}\mathbf{E}^*$ and $\mathbf{C} = \mathbf{R}_{\mathbf{yx}}\mathbf{E} + \mathbf{P}_{\mathbf{yx}}\mathbf{D}^*$ are coefficient matrices, with $\mathbf{D} = (\mathbf{R}_{\mathbf{x}} - \mathbf{P}_{\mathbf{x}}\mathbf{R}_{\mathbf{x}}^{*-1}\mathbf{P}_{\mathbf{x}}^*)^{-1}$ and $\mathbf{E} = -(\mathbf{R}_{\mathbf{x}} - \mathbf{P}_{\mathbf{x}}\mathbf{R}_{\mathbf{x}}^{*-1}\mathbf{P}_{\mathbf{x}}^*)^{-1}\mathbf{P}_{\mathbf{x}}\mathbf{R}_{\mathbf{x}}^{*-1}$, while $\mathbf{x}^a = [\mathbf{x}^T, \mathbf{x}^H]^T$ is the augmented input vector, and $\mathbf{W} = [\mathbf{B}, \mathbf{C}]$ the optimal coefficient matrix. The estimator in \eqref{Eq:WL_model} is optimal for the generality of complex signals, both circular and noncircular. The full second order information is contained in the augmented covariance matrix 
\begin{eqnarray}
\label{rza1}
	\mathbf{R}^a_{\mathbf{x}} = E\{\mathbf{x}^a\mathbf{x}^{aH}\} = \begin{bmatrix} \mathbf{R}_{\mathbf{x}} & \mathbf{P}_{\mathbf{x}}\\\mathbf{P}^*_{\mathbf{x}} & \mathbf{R}^*_{\mathbf{\mathbf{x}}} \end{bmatrix}
\end{eqnarray}
and as such, estimation based on $\mathbf{R}^a_{\mathbf{x}}$ incorporates both the covariance and pseudocovariance, and applies to both proper and improper data \cite{Picinbono97, Dini_Class_WLKF_IEEE_TNNLS_2012, Moreno_2009}.
%

Extending the recent work on widely linear estimation and distributed Kalman filters \cite{Mandic09,Dini_Class_WLKF_IEEE_TNNLS_2012, Olfati_Saber_Dist_KF, Cattivelli_Sayed_IEEETranAC_Dist_KF_2010}, we here propose a distributed augmented (widely linear) complex Kalman filter (D-ACKF) that caters for general complex signals, as well as the cross-correlations between the observation noises at neighbouring nodes. These are real-world scenarios encountered when node signals are exposed to common noise, such as in multi-sensor target tracking in the presence of observation jamming-noise, environmental noise in seismic arrays, signals from microphone array systems experiencing common interference, wireless sensor networks with overlapping user frequencies, and distributed frequency estimation in smart grids experiencing common fault.

This work generalises earlier distributed Kalman filtering approaches \cite{Olfati_Saber_Dist_KF, Kar_Moura_Dist_KF_IEEETransSP_2011, Cattivelli_Sayed_IEEETranAC_Dist_KF_2010} and illuminates the duality of D-ACKF with its corresponding bivariate real-valued distributed Kalman filter, highlighting several issues of implementation motivated by duality considerations. The performance of the D-ACKF is analysed, and supported by case studies on filtering autoregressive processes and projectile tracking, involving both proper and improper signals.
%
\section{Diffusion Kalman Filtering}
Consider the standard linear state space corresponding to a node $i$ in a distributed system \cite{Hayes96},
\begin{subequations}\label{linearstatespace}
\begin{eqnarray}
	\mathbf{x}_{n} &=& \mathbf{F}_{n-1}\mathbf{x}_{n-1} + \mathbf{w}_{n} \label{s1}\\
	\mathbf{y}_{i,n} &=& \mathbf{H}_{i,n}\mathbf{x}_{n} + \mathbf{v}_{i,n} \label{dm1}
\end{eqnarray}
\end{subequations}
where $\mathbf{x}_{n} \in \mathbb{C}^L$ and $\mathbf{y}_{i,n} \in \mathbb{C}^{K}$ are respectively the state vector at time instant $n$ and observation (measurement) vector at node $i$, while $\mathbf{F}_n$ and $\mathbf{H}_{i,n}$ are the state transition and observation matrices, whereas $\mathbf{w}_{n} \in \mathbb{C}^L$ and $\mathbf{v}_{i,n} \in \mathbb{C}^{K}$ are respectively the white state and measurement noises at node $i$, and are assumed to be uncorrelated and zero-mean, with covariances and pseudocovariances defined as
\begin{eqnarray}
	E \begin{bmatrix} \mathbf{w}_{n} \\ \mathbf{v}_{i,n} \end{bmatrix} \begin{bmatrix} \mathbf{w}_{k} \\ \mathbf{v}_{i,k} \end{bmatrix}^H
	=
	\begin{bmatrix} \mathbf{Q}_n & \mathbf{0} \\ \mathbf{0} & \mathbf{R}_{i,n} \end{bmatrix} \delta_{nk} \\
	E \begin{bmatrix} \mathbf{w}_{n} \\ \mathbf{v}_{i,n} \end{bmatrix} \begin{bmatrix} \mathbf{w}_{k} \\ \mathbf{v}_{i,k} \end{bmatrix}^T
	=
	\begin{bmatrix} \mathbf{P}_n & \mathbf{0} \\ \mathbf{0} & \mathbf{U}_{i,n} \end{bmatrix} \delta_{nk}
\end{eqnarray}
where $\delta_{nk}$ is the Kronecker delta function. 
%
\subsection{Distributed Complex Kalman Filter}\label{Sec:D-CKF}
The distinguishing feature of the proposed class of distributed Kalman filters is that no assumption is made about the correlation of the observation noises at different nodes, thus extending earlier distributed Kalman filtering algorithms  \cite{Cattivelli_Sayed_IEEETranAC_Dist_KF_2010, Olfati_Saber_Dist_KF, Kar_Moura_Dist_KF_IEEETransSP_2011}, and allowing us to deal more effectively with cases where the nodes experience common measurement noises.

\begin{figure}[t]
  \centering  
  \includegraphics[trim = 4mm 5mm 4mm 15mm,scale=0.6] {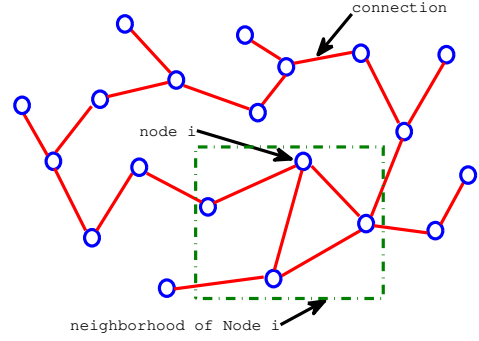}
  \caption{An illustrative example of a distributed network topology.}
  \label{fig:NetwotkTopologyGeneral}
\end{figure}

Denote the neighbourhood of node $i$, that is, the set of nodes that can communicate directly with the node $i$ (including itself) by $\mathcal{N}_i$, as illustrated in Figure \ref{fig:NetwotkTopologyGeneral}. 

Let $\mathbf{\widehat{\underline{x}}}_{i,n|n}$ denote the  complex Kalman filter (CKF) state estimate at node $i$ based on all the data from the neighbourhood $\mathcal{N}_i$ consisting of $M = |\mathcal{N}_i|$ nodes, where $|\mathcal{N}_i|$ denotes the number of nodes in the neighbourhood $\mathcal{N}_i$. The collective neighbourhood observation equation at node $i$ is given by
\begin{eqnarray} \label{c_neighbourhood_obs_eqn}
	\mathbf{\underline{y}}_{i,n} = \mathbf{\underline{H}}_{i,n}\mathbf{x}_{n} + \mathbf{\underline{v}}_{i,n}
\end{eqnarray}
with the collective (neighbourhood)) variables defined as
\begin{align*}
	\mathbf{\underline{y}}_{i,n} &= \begin{bmatrix} \mathbf{y}_{i_1,n}^T, \mathbf{y}_{i_2,n}^T, \cdots , \mathbf{y}_{i_M,n}^T \end{bmatrix}^T\\ 
	\mathbf{\underline{H}}_{i,n} &= \begin{bmatrix} \mathbf{H}_{i_1,n}^T, \mathbf{H}_{i_2,n}^T, \cdots , \mathbf{H}_{i_M,n}^T \end{bmatrix}^T\\
	\mathbf{\underline{v}}_{i,n} &= \begin{bmatrix} \mathbf{v}_{i_1,n}^T, \mathbf{v}_{i_2,n}^T, \cdots , \mathbf{v}_{i_M,n}^T \end{bmatrix}^T
\end{align*}
where $\{i_1, i_2, \ldots, i_M\}$ are all the nodes in the neighbourhood $\mathcal{N}_i$. The covariance and pseudocovariance of the collective observation noise vector are:
\begin{align*}
	&\mathbf{\underline{R}}_{i,n} = E\{\mathbf{\underline{v}}_{i,n}\mathbf{\underline{v}}_{i,n}^H \} 
	= 
	\begin{bmatrix}[l] \mathbf{R}_{i_1,n}    	& \mathbf{R}_{i_1i_2,n} & \cdots & \mathbf{R}_{i_1i_M,n}  \\
	    						\mathbf{R}_{i_2i_1,n} 	& \mathbf{R}_{i_2,n}    & \cdots & \mathbf{R}_{i_2i_M,n}  \\
	    						\vdots 									& \vdots 								& \ddots & \vdots 								\\
	    						\mathbf{R}_{i_Mi_1,n}  	& \mathbf{R}_{i_Mi_2,n} & \cdots & \mathbf{R}_{i_M,n}				
	\end{bmatrix} \\
	&\mathbf{\underline{U}}_{i,n} = E\{\mathbf{\underline{v}}_{i,n}\mathbf{\underline{v}}_{i,n}^T \}
	= 
	\begin{bmatrix}[l] \mathbf{U}_{i_1,n}    	& \mathbf{U}_{i_1i_2,n} & \cdots & \mathbf{U}_{i_1i_M,n}  \\
	    						\mathbf{U}_{i_2i_1,n} 	& \mathbf{U}_{i_2,n}    & \cdots & \mathbf{U}_{i_2i_M,n}  \\
	    						\vdots 									& \vdots 								& \ddots & \vdots 								\\
	    						\mathbf{U}_{i_Mi_1,n}  	& \mathbf{U}_{i_Mi_2,n} & \cdots & \mathbf{U}_{i_M,n}				
	\end{bmatrix} 	
\end{align*}
where $\mathbf{R}_{i_a,n} = E\{\mathbf{v}_{i_a,n}\mathbf{v}_{i_a,n}^H\}$, $\mathbf{R}_{i_ai_b,n} = E\{\mathbf{v}_{i_a,n}\mathbf{v}_{i_b,n}^H\}$, $\mathbf{U}_{i_a,n} = E\{\mathbf{v}_{i_a,n}\mathbf{v}_{i_a,n}^T\}$ and $\mathbf{U}_{i_ai_b,n} = E\{\mathbf{v}_{i_a,n}\mathbf{v}_{i_b,n}^T\}$, for $a,b \in \{1,2,\ldots , M\}$. 

Calculation of the neighbourhood state estimates is followed by the diffusion step, given by
\begin{eqnarray}
	\mathbf{\widehat{x}}_{i,n|n} = \sum_{k \in \mathcal{N}_i} c_{k,i}\mathbf{\widehat{\underline{x}}}_{k,n|n}
\end{eqnarray}
where the diffused state estimates $\mathbf{\widehat{x}}_{i,n|n}$ is the weighted estimates from the neighbourhood $\mathcal{N}_i$, and $c_{k,i} \geq 0$ are the weighting coefficients satisfying $\sum_{k \in \mathcal{N}_i} c_{k,i} = 1$. 
A number of fusion schemes have been proposed, including the Metropolis \cite{Lopes_Sayed_TSP_Dist_LMS_2008}, Laplacian \cite{Bru_Convergence_1994} and nearest neighbour method \cite{Cattivelli_Sayed_IEEETranAC_Dist_KF_2010}, however, the determination of the optimal weights for an arbitrary network of nodes is a difficult problem without accurate knowledge of the statistics of the local estimates \cite{Li_Optimal_Dist_fusion_2003}. 

The distributed complex Kalman filter (D-CKF) aims to approximate a centralised Kalman filter (with access to the observation data from all the nodes) via neighbourhood collaborations and diffusion, and is summarised in Algorithm \ref{alg:D-CKF}. The D-CKF algorithm requires each node to form a collective observation equation as in \eqref{c_neighbourhood_obs_eqn} by gathering information from its neighbours, thereafter, each node computes a neighbourhood state estimate which are again transmitted to neighbours to be used for the diffusion step. 
%
\begin{algorithm}[t]
Initialisation: For each node $i=1,2,\ldots,N$
\begin{eqnarray}
	\mathbf{\widehat{x}}_{i,0|0} &=& E\{\mathbf{x}_{0}\} \nonumber\\
	\mathbf{M}_{i,0|0} &=& E\{(\mathbf{x}_{0}-E\{\mathbf{x}_{0}\})(\mathbf{x}_{0}-E\{\mathbf{x}_{0}\})^H\} \nonumber
\end{eqnarray} 
For every time instant $n=1,2,\ldots$ 

$-$ Evaluate at each node $i=1,2,\ldots,N$
\begin{align}
	\mathbf{\widehat{x}}_{i,n|n-1} &= \mathbf{F}_{n-1}\mathbf{\widehat{x}}_{i,n-1|n-1} \label{px_est}\\
  \mathbf{M}_{i,n|n-1} &= \mathbf{F}_{n-1}\mathbf{M}_{i,n-1|n-1}\mathbf{F}^{H}_{n-1} + \mathbf{Q}_n \label{pMSE}\\
	\mathbf{G}_{i,n} &= \mathbf{M}_{i,n|n-1}\mathbf{\underline{H}}^{H}_{i,n}\big(\mathbf{\underline{H}}_{i,n}\mathbf{M}_{i,n|n-1}\mathbf{\underline{H}}^{H}_{i,n} + \mathbf{\underline{R}}_{i,n}\big)^{-1} \label{gain}\\
	\mathbf{\widehat{\underline{x}}}_{i,n|n} &= \mathbf{\widehat{x}}_{i,n|n-1} + \mathbf{G}_{i,n}\big(\mathbf{\underline{y}}_{i,n} - \mathbf{\underline{H}}_{i,n}\mathbf{\widehat{x}}_{i,n|n-1}\big)\label{x_est}\\
	\mathbf{M}_{i,n|n} &= (\mathbf{I} - \mathbf{G}_{i,n}\mathbf{\underline{H}}_{i,n})\mathbf{M}_{i,n|n-1}\label{MSE}
\end{align}
$-$ For every node $i$, compute the diffusion update as
\begin{align}
	\mathbf{\widehat{x}}_{i,n|n} = \sum_{k \in \mathcal{N}_i} c_{k,i}\mathbf{\widehat{\underline{x}}}_{k,n|n}
\end{align}
\caption{The D-CKF}
\label{alg:D-CKF}
\end{algorithm}
\newline
\textbf{Remark \#1:} The D-CKF algorithm\footnote{The matrices $\mathbf{M}_{i,n|n}$ and $\mathbf{M}_{i,n|n-1}$ do not represent the covariances of $\mathbf{\widehat{x}}_{i,n|n}$ and $\mathbf{\widehat{x}}_{i,n|n-1}$, as is the case for the standard Kalman filter operating on linear Gaussian systems. This is due to the use of the suboptimal diffusion step, which updates the state estimate and not the covariance matrix $\mathbf{M}_{i,n|n}$.} is based on the standard (strictly linear) state space model \eqref{linearstatespace}, similar to existing algorithms \cite{Cattivelli_Sayed_IEEETranAC_Dist_KF_2010} \cite{Cattivelli_Sayed_ICASSP_Dist_KF_2010}, and is thus inadequate for widely linear state space models or noncircular state and observation noises, where $\mathbf{P}_{n} \neq \mathbf{0}$ and $\mathbf{U}_{i,n} \neq \mathbf{0}$ for $i=1,2,\ldots,N$. 
\newline\newline
\textbf{Remark \#2:} Observe that unlike existing distributed complex Kalman filters, the proposed D-CKF algorithm also caters for the cross-correlations between the neighbourhood observation noises, while for uncorrelated nodal observation noises, it degenerates into Algorithm \ref{alg:D-CKF} in \cite{Cattivelli_Sayed_IEEETranAC_Dist_KF_2010}.
%
%
\subsection{Distributed Augmented Complex Kalman Filter} \label{Sec:D-ACKF}
To cater for widely linear state and observation models together with improper signals, based the widely linear model in \eqref{Eq:WL_model}, the widely linear version of the distributed state space model \eqref{linearstatespace} is defined as
\cite{Dini_Class_WLKF_IEEE_TNNLS_2012}
\begin{subequations}\label{WLstatespace}
\begin{eqnarray}
	\mathbf{x}_{n} &=& \mathbf{F}_{n-1}\mathbf{x}_{n-1} + \mathbf{A}_{n-1}\mathbf{x}^*_{n-1} + \mathbf{w}_{n} \\
	\mathbf{y}_{i,n} &=& \mathbf{H}_{i,n}\mathbf{x}_{n} + \mathbf{B}_{i,n}\mathbf{x}_{n}^* + \mathbf{v}_{i,n} \label{WLdm1}
\end{eqnarray}  
\end{subequations}
or in its augmented representation:
\begin{subequations}\label{AUGstatespace}
\begin{eqnarray}
	\mathbf{x}^a_{n} &=& \mathbf{F}^a_{n-1}\mathbf{x}^a_{n-1} + \mathbf{w}^a_{n} \label{as1}\\
	\mathbf{y}_{i,n}^a &=& \mathbf{H}_{i,n}^a\mathbf{x}_{n}^a + \mathbf{v}_{i,n}^a 
\end{eqnarray} 
\end{subequations}  
where $\mathbf{x}^a_n = [\mathbf{x}^T_n, \mathbf{x}^H_n]^T$ and $\mathbf{y}^a_n = [\mathbf{y}^T_n, \mathbf{y}^H_n]^T$, while
\begin{eqnarray*}
	\mathbf{F}^a_{n} = \begin{bmatrix} \mathbf{F}_{n} & \mathbf{A}_{n}\\ \mathbf{A}^*_{n} & \mathbf{F}^*_{n} \end{bmatrix}
	\text{  and  }
	\mathbf{H}^a_{i,n} = \begin{bmatrix} \mathbf{H}_{i,n} & \mathbf{B}_{i,n}\\ \mathbf{B}^*_{i,n} & \mathbf{H}^*_{i,n} \end{bmatrix}
\end{eqnarray*} 			
\newline\newline
\textbf{Remark \#3:} For strictly linear systems, $\mathbf{A}_{n} = \mathbf{0}$ and $\mathbf{B}_{i,n} = \mathbf{0}$, so that the widely linear (augmented) state space model degenerates into a strictly linear one, however, the augmented state space representation is still preferred in order to account for the pseudocovariances (impropriety) of the signals (\textit{cf.} widely linear systems). 

The augmented covariance matrices of $\mathbf{w}^a_n = [\mathbf{x}^T_n, \mathbf{w}^H_n]^T$ and $\mathbf{v}^a_{i,n} = [\mathbf{v}^T_{i,n}, \mathbf{v}^H_{i,n}]^T$ are then given by
\begin{eqnarray}
	\mathbf{Q}^a_n &=& E\{\mathbf{w}^a_{n}\mathbf{w}^{aH}_{n}\} =\begin{bmatrix} \mathbf{Q}_n & \mathbf{P}_n\\ \mathbf{P}^*_n & \mathbf{Q}^*_n \end{bmatrix} \label{aQ}\\
	\mathbf{R}^a_{i,n} &=& E\{\mathbf{v}^a_{i,n}\mathbf{v}^{aH}_{i,n}\} =\begin{bmatrix} \mathbf{R}_{i,n} & \mathbf{U}_{i,n}\\ \mathbf{U}^*_{i,n} & \mathbf{R}^*_{i,n} \end{bmatrix}\label{aR}	
\end{eqnarray}

To enable collaborative estimation of the state within distributed networks, we employ neighbourhood observation equations comprising of all the neighbourhood observation data, that is
\begin{eqnarray} \label{ac_neighbourhood_obs_eqn}
	\mathbf{\underline{y}}_{i,n} = \mathbf{\underline{H}}_{i,n}\mathbf{x}_{n} + \mathbf{\underline{B}}_{i,n}\mathbf{x}_{n}^* +\mathbf{\underline{v}}_{i,n}
\end{eqnarray}
where the conjugate state matrix $\mathbf{\underline{B}}_{i,n} = \big[\mathbf{B}_{i_1,n}^T, \mathbf{B}_{i_2,n}^T, \ldots, \mathbf{B}_{i_M,n}^T \big]^T$, and $\{i_1, i_2, \ldots, i_M\} \in \mathcal{N}_i$. The augmented neighbourhood observation equations can now be written as
\begin{eqnarray}\label{aug_y_collective}
	\mathbf{\underline{y}}_{i,n}^a = \mathbf{\underline{H}}_{i,n}^a\mathbf{x}_{n}^a + \mathbf{\underline{v}}_{i,n}^a 
\end{eqnarray}
with augmented neighbourhood terms defined as
\begin{eqnarray}
	\mathbf{\underline{y}}_{i,n}^a = \begin{bmatrix} \mathbf{\underline{y}}_{i,n} \\ \mathbf{\underline{y}}_{i,n}^* \end{bmatrix}, 
	\quad
	\mathbf{\underline{H}}_{i,n}^a = \begin{bmatrix} \mathbf{\underline{H}}_{i,n} & \mathbf{\underline{B}}_{i,n}\\ 
																	 \mathbf{\underline{B}}_{i,n}^* & \mathbf{\underline{H}}_{i,n}^*\end{bmatrix},
	\quad
	\mathbf{\underline{v}}_{i,n}^a = \begin{bmatrix} \mathbf{\underline{v}}_{i,n} \\ \mathbf{\underline{v}}_{i,n}^* \end{bmatrix}
\end{eqnarray}
The covariance of the augmented noise $\mathbf{\underline{v}}_{i,n}^a$ is then defined as
\begin{eqnarray}\label{aug_R_collective}
	\mathbf{\underline{R}}_{i,n}^a 
	= E\{\mathbf{\underline{v}}^a_{i,n}\mathbf{\underline{v}}^{aH}_{i,n}\} 
	= \begin{bmatrix} \mathbf{\underline{R}}_{i,n} & \mathbf{\underline{U}}_{i,n}\\ \mathbf{\underline{U}}^*_{i,n} & \mathbf{\underline{R}}^*_{i,n} \end{bmatrix}	
\end{eqnarray}
and caters for both the covariances $E\{\mathbf{v}_{i,n}\mathbf{v}_{i,n}^H\}$ and cross-correlations $E\{\mathbf{v}_{i,n}\mathbf{v}_{k,n}^H\}$, $i\neq k$ of the nodal observation noises through the covariance matrix $\mathbf{\underline{R}}_{i,n}$, and the pseudocovariances $E\{\mathbf{v}_{i,n}\mathbf{v}_{i,n}^T\}$ and cross-pseudocorrelations $E\{\mathbf{v}_{i,n}\mathbf{v}_{k,n}^T\}$ through the pseudocovariance matrix $\mathbf{\underline{U}}_{i,n}$.
Finally, the augmented diffused state estimate becomes
\begin{eqnarray}
	\mathbf{\widehat{x}}_{i,n|n}^a = \sum_{k \in \mathcal{N}_i} c_{k,i}\mathbf{\widehat{\underline{x}}}_{k,n|n}^a
\end{eqnarray}
and represents a weighted average of the augmented (neighbourhood) state estimates. The proposed distributed augmented complex Kalman filter (D-ACKF), based on the widely linear state space model, is summarised in Algorithm \ref{alg:D-ACKF}. 
\newline\newline
\textbf{Remark \#4:} For strictly linear systems ($\mathbf{A}_{n} = \mathbf{0}$ and $\mathbf{B}_{i,n} = \mathbf{0}$ for all $n$ and $i$) with circular state and observation noises ($\mathbf{P}_{n} = \mathbf{0}$ and $\mathbf{U}_{i,n} = \mathbf{0}$ for all $n$ and $i$), the D-ACKF and D-CKF algorithms are equivalent, in the sense that they yield identical state estimates for all time instants $n$. 

However, if any of these conditions are not met, the D-ACKF assumes a more general form than the D-CKF. This can be illustrated based on the analysis in \cite{Dini_Class_WLKF_IEEE_TNNLS_2012}, which shows the advantages of the augmented complex Kalman filter (ACKF) over the conventional (strictly linear) complex Kalman filter (CKF) for the non-distributed case. 
\begin{algorithm}[t]
Initialisation: For each node $i=1,2,\ldots,N$
\begin{eqnarray*}
	\mathbf{\widehat{x}}_{i,0|0}^a &=& \big[E\{\mathbf{x}_{0}\}^T, E\{\mathbf{x}_{0}\}^H\big]^T \\
	\mathbf{M}_{i,0|0}^a &=& E\big\{(\mathbf{x}_{0}^a-\mathbf{\widehat{x}}_{i,0|0}^a)(\mathbf{x}_{0}^a-\mathbf{\widehat{x}}_{i,0|0}^a)^{aH}\big\}
\end{eqnarray*} 
For every time instant $n=1,2,\ldots$ 

\hspace{0.15cm}$-$ Evaluate at each node $i=1,2,\ldots,N$
\begin{align}
	\mathbf{\widehat{x}}_{i,n|n-1}^a &= \mathbf{F}_{n-1}^a\mathbf{\widehat{x}}_{i,n-1|n-1}^a \label{augpx_est}\\
  \mathbf{M}_{i,n|n-1}^a &= \mathbf{F}_{n-1}^a\mathbf{M}_{i,n-1|n-1}^a\mathbf{F}^{aH}_{n-1} + \mathbf{Q}_n^a \label{augpMSE}\\
	\mathbf{G}_{i,n}^a &= \mathbf{M}_{i,n|n-1}^a\mathbf{\underline{H}}^{aH}_{i,n}\big(\mathbf{\underline{H}}_{i,n}^a\mathbf{M}_{i,n|n-1}^a\mathbf{\underline{H}}^{aH}_{i,n} + \mathbf{\underline{R}}_{i,n}^a\big)^{-1} \label{aug_gain}\\
	\mathbf{\widehat{\underline{x}}}_{i,n|n}^a &= \mathbf{\widehat{x}}_{i,n|n-1}^a + \mathbf{G}_{i,n}^a\big(\mathbf{\underline{y}}_{i,n}^a - \mathbf{\underline{H}}_{i,n}^a\mathbf{\widehat{x}}_{i,n|n-1}^a\big)\label{augx_est}\\
	\mathbf{M}_{i,n|n}^a &= (\mathbf{I} - \mathbf{G}_{i,n}^a\mathbf{\underline{H}}_{i,n}^a)\mathbf{M}_{i,n|n-1}^a\label{augMSE}
\end{align}
\hspace{0.15cm}$-$ For every node $i$, compute the diffusion update as
\begin{align}
	\mathbf{\widehat{x}}_{i,n|n}^a = {\sum}_{k \in \mathcal{N}_i} c_{k,i}\mathbf{\widehat{\underline{x}}}_{k,n|n}^a \label{augdx}
\end{align}
\caption{The D-ACKF}
\label{alg:D-ACKF}
\end{algorithm}
\newline\newline
\textbf{Remark \#5:} When the nodes are subject to uncorrelated observation noises, the information form of the D-ACKF, given in Algorithm \ref{alg:D-ACKF_info}, can be utilised to cater for the propriety of the signals without accounting for observation noise correlations at different nodes. 

Further, depending on the correlation between the observation noises, different nodes in the distributed network can switch between the general D-ACKF in Algorithm \ref{alg:D-ACKF} and the information form D-ACKF in Algorithm \ref{alg:D-ACKF_info}.
%
%
%
\section{Analysis}
\subsection{Duality Analysis}
Owing to the isomorphism between augmented complex vectors and bivariate real vectors, and the duality analysis for stochastic gradient filters \cite{Mandic_Duality_ICASSP_2009}, the D-ACKF algorithm has a dual bivariate distributed real valued Kalman filter (D-RKF) which can be used to reduce its computational complexity. 

A complex vector $\mathbf{z} = \mathbf{z}_r + j\mathbf{z}_i \in \mathbb{C}^q$ has a composite bivariate real representation in $\mathbb{R}^{2q}$ of the form 
\begin{eqnarray}\label{duality_C_R}
	\mathbf{z}^a = \begin{bmatrix} \mathbf{z} \\ \mathbf{z}^* \end{bmatrix} = \underbrace{\begin{bmatrix} \mathbf{I} & j\mathbf{I} \\ \mathbf{I} & -j\mathbf{I} \end{bmatrix}}_{\equiv\mathbf{J}_{\mathbf{z}}} \underbrace{\begin{bmatrix} \mathbf{\mathbf{z}_r} \\ \mathbf{\mathbf{z}_i} \end{bmatrix}}_{=\mathbf{z}^r}
\end{eqnarray}
where $\mathbf{I}$ is the identity matrix (with appropriate dimensions), and the invertible orthogonal mapping\footnote{For a vector $\mathbf{z} \in \mathbb{C}^q$, the corresponding orthogonal matrix $\mathbf{J}_{\mathbf{z}}$ takes dimension $2q\times 2q$.} $\mathbf{J}_{\mathbf{z}}:\mathbb{C}^{2q}\rightarrow\mathbb{R}^{2q}$ is such that $\mathbf{J}_{\mathbf{z}}^{-1}=\frac{1}{2}\mathbf{J}_{\mathbf{z}}^H$ \cite{Brandwood_Complex_Grad_1983}\cite{Delgado06}. Based on this isomorphism, the real bivariate state space corresponding to the augmented complex state space in \eqref{AUGstatespace} is given by
\begin{subequations}
\begin{eqnarray}
	\mathbf{x}^r_{n} &=& \mathbf{F}^r_{n-1}\mathbf{x}^r_{n-1} + \mathbf{w}^r_{n} \nonumber\\
	\mathbf{y}^r_{n} &=& \mathbf{H}^r_{n}\mathbf{x}^r_{n} + \mathbf{v}^r_{n}\label{rs1}
\end{eqnarray} 
\end{subequations}
\begin{algorithm}[t]
Initialisation: For each node $i=1,2,\ldots,N$
\begin{eqnarray*}
	\mathbf{\widehat{x}}_{i,0|0}^a &=& \big[E\{\mathbf{x}_{0}\}^T, E\{\mathbf{x}_{0}\}^H\big]^T \\
	\mathbf{M}_{i,0|0}^a &=& E\big\{(\mathbf{x}_{0}^a-\mathbf{\widehat{x}}_{i,0|0}^a)(\mathbf{x}_{0}^a-\mathbf{\widehat{x}}_{i,0|0}^a)^{aH}\big\}
\end{eqnarray*} 
For every time instant $n=1,2,\ldots$ 

\hspace{0.15cm}$-$ Evaluate at each node $i=1,2,\ldots,N$
\begin{align}
	\mathbf{\widehat{x}}_{i,n|n-1}^a &= \mathbf{F}_{n-1}^a\mathbf{\widehat{x}}_{i,n-1|n-1}^a \label{augpx_est_info}\\
  \mathbf{M}_{i,n|n-1}^a &= \mathbf{F}_{n-1}^a\mathbf{M}_{i,n-1|n-1}^a\mathbf{F}^{aH}_{n-1} + \mathbf{Q}_n^a \label{augpMSE_info}\\
	\mathbf{S}_{i,n}^a &= {\sum}_{k \in \mathcal{N}_i} \mathbf{H}^{aH}_{k,n}(\mathbf{R}_{k,n}^a)^{-1}\mathbf{H}_{k,n}^a \\
	\mathbf{r}_{i,n}^a &= {\sum}_{k \in \mathcal{N}_i} \mathbf{H}^{aH}_{k,n}(\mathbf{R}_{k,n}^a)^{-1}\mathbf{y}_{k,n}^a \\
	(\mathbf{M}_{i,n|n}^a)^{-1} &= (\mathbf{M}_{i,n|n-1}^a)^{-1} + \mathbf{S}_{i,n}^a\\
	\mathbf{\widehat{\chi}}_{i,n|n}^a &= \mathbf{\widehat{x}}_{i,n|n-1}^a + \mathbf{M}_{i,n|n}^a\big(\mathbf{r}_{i,n}^a - \mathbf{S}_{i,n}^a\mathbf{\widehat{x}}_{i,n|n-1}^a\big)
\end{align}
\hspace{0.15cm}$-$ For every node $i$, compute the diffusion update as
\begin{align}
	\mathbf{\widehat{x}}_{i,n|n}^a = {\sum}_{k \in \mathcal{N}_i} c_{k,i}\mathbf{\widehat{\chi}}_{i,n|n}^a \label{augdx_info}
\end{align}
\caption{The D-ACKF Information Form}
\label{alg:D-ACKF_info}
\end{algorithm}
where $\mathbf{x}^r_{n}= \mathbf{J}^{-1}_{\mathbf{x}}\mathbf{x}^a_{n}$, $\mathbf{y}^r_{n}=\mathbf{J}^{-1}_{\mathbf{y}}\mathbf{y}^a_{n}$, $\mathbf{F}^r_{n-1} = \mathbf{J}^{-1}_{\mathbf{x}}\mathbf{F}^a_{n-1}\mathbf{J}_{\mathbf{x}}$, $\mathbf{H}^r_{n}=\mathbf{J}_{\mathbf{y}}^{-1}\mathbf{H}^a_{n}\mathbf{J}_{\mathbf{x}}$, $\mathbf{w}^r_{n}=\mathbf{J}_{\mathbf{x}}^{-1}\mathbf{w}^a_{n}$ and $\mathbf{v}^r_{n}=\mathbf{J}_{\mathbf{y}}^{-1}\mathbf{v}^a_{n}$. In the same vein, the real valued covariance matrices of $\mathbf{w}^r_{n}$ and $\mathbf{v}^r_{n}$ take the corresponding forms
\begin{eqnarray}
	\mathbf{Q}^r_{n} &=& E\{\mathbf{w}^r_{n}\mathbf{w}^{rH}_{n}\} = \mathbf{J}_{\mathbf{x}}^{-1}\mathbf{Q}^a_{n}\mathbf{J}_{\mathbf{x}}^{-H} \nonumber\\
	\mathbf{R}^r_{n} &=& E\{\mathbf{v}^r_{n}\mathbf{v}^{rH}_{n}\} = \mathbf{J}_{\mathbf{y}}^{-1}\mathbf{R}^a_{n}\mathbf{J}_{\mathbf{y}}^{-H} \nonumber
\end{eqnarray} 
while the real valued counterpart of \eqref{aug_y_collective} is given by
\begin{eqnarray}
	\mathbf{\underline{y}}_{i,n}^r = \mathbf{\underline{H}}_{i,n}^r\mathbf{x}_{n}^r + \mathbf{\underline{v}}_{i,n}^r 
\end{eqnarray}
with $\mathbf{\underline{y}}^r_{n}=\mathbf{J}^{-1}_{\mathbf{\underline{y}}}\mathbf{\underline{y}}^a_{n}$,
$\mathbf{\underline{H}}^r_{n}=\mathbf{J}_{\mathbf{\underline{y}}}^{-1}\mathbf{\underline{H}}^a_{n}\mathbf{J}_{\mathbf{x}}$ and
$\mathbf{\underline{v}}^r_{n}=\mathbf{J}_{\mathbf{\underline{y}}}^{-1}\mathbf{\underline{v}}^a_{n}$. Finally, the covariance matrix of $\mathbf{\underline{v}}^r_{n}$ is defined as
\begin{eqnarray*}
	\mathbf{\underline{R}}^r_{n} &=& E\{\mathbf{\underline{v}}^r_{n}\mathbf{\underline{v}}^{rH}_{n}\} = 
																	 \mathbf{J}_{\mathbf{\underline{y}}}^{-1}\mathbf{\underline{R}}^a_{n}\mathbf{J}_{\mathbf{\underline{y}}}^{-H} 
\end{eqnarray*} 
The duality between the D-ACKF and the D-RKF is established through the following relationships:
\begin{eqnarray*}\label{D-ACKF<->ARDKF_equivelence}
	\mathbf{\widehat{x}}^r_{i,n|n-1} &=& \mathbf{J}_{\mathbf{x}}^{-1}\mathbf{\widehat{x}}^a_{i,n|n-1} \\
	\mathbf{M}^r_{i,n|n-1} &=& \mathbf{J}_{\mathbf{x}}^{-1}\mathbf{M}^a_{i,n|n-1}\mathbf{J}_{\mathbf{x}}^{-H}\\
	\mathbf{G}^r_{i,n} &=& \mathbf{J}_{\mathbf{x}}^{-1}\mathbf{G}^a_{i,n}\mathbf{J}_{\mathbf{\underline{y}}}\\
	\mathbf{\widehat{\underline{x}}}^r_{i,n|n} &=& \mathbf{J}_{\mathbf{x}}^{-1}\mathbf{\widehat{\underline{x}}}^a_{i,n|n}\\
	\mathbf{M}^r_{i,n|n} &=& \mathbf{J}_{\mathbf{x}}^{-1}\mathbf{M}^a_{i,n|n}\mathbf{J}_{\mathbf{x}}^{-H}\\
	\mathbf{\widehat{x}}^r_{i,n|n} &=& \mathbf{J}_{\mathbf{x}}^{-1}\mathbf{\widehat{x}}^a_{i,n|n}
\end{eqnarray*}

Therefore, the D-ACKF and D-RKF effectively implement the same state space model, but operate in the complex and real domains, respectively. Generally speaking, for systems naturally defined in the complex domain, it is desirable to keep the computations in the original complex domain in order to facilitate understanding of the signal transformations, together with benefiting from the well defined notions of phase and circularity.

However, since the computation cost associated with complex-valued algorithms is higher than that for real valued ones, the duality analysis provides a framework for reducing these cost, whereby the number of additions and multiplications required are approximately halved and quartered, respectively.
%
%
\subsection{Mean And Mean Square Analysis}
%
%
For generality in analysis, we consider augmented complex variables: let $\mathbf{\underline{e}}_{i,n|n}^a = \mathbf{x}_{n}^a - \mathbf{\widehat{\underline{x}}}_{i,n|n}^a$ denote the local (non-diffused) error at node $i \in [1,N]$, $\mathbf{e}_{i,n|n-1}^a = \mathbf{x}_{n}^a - \mathbf{\widehat{x}}_{i,n|n-1}^a$ the prediction error, and $\mathbf{e}_{i,n|n}^a = \mathbf{x}_{n}^a - \mathbf{\widehat{x}}_{i,n|n}^a$ the diffused error. The difference between the true state in \eqref{as1} and the predicted state estimate in \eqref{augx_est} now becomes
\begin{align}\label{augLocalError}
	\mathbf{\underline{e}}_{i,n|n}^a 	&= \mathbf{x}_{n}^a - \mathbf{\widehat{x}}_{i,n|n-1}^a + \mathbf{G}_{i,n}^a\big(\mathbf{\underline{y}}_{i,n}^a -\mathbf{\underline{H}}_{i,n}^a\mathbf{\widehat{x}}_{i,n|n-1}^a\big)\nonumber\\
	&= \mathbf{e}_{i,n|n-1}^a + \mathbf{G}_{i,n}^a\big(\mathbf{\underline{H}}_{i,n}^a\mathbf{x}_{n}^a + \mathbf{\underline{v}}_{i,n}^a -\mathbf{\underline{H}}_{i,n}^a\mathbf{\widehat{x}}_{i,n|n-1}^a\big)\nonumber\\
	&= \mathbf{e}_{i,n|n-1}^a + \mathbf{G}_{i,n}^a\big(\mathbf{\underline{H}}_{i,n}^a\mathbf{e}_{i,n|n-1}^a + \mathbf{\underline{v}}_{i,n}^a \big)\nonumber\\
	&= \big(\mathbf{I} + \mathbf{G}_{i,n}^a\mathbf{\underline{H}}_{i,n}^a\big)\mathbf{e}_{i,n|n-1}^a + \mathbf{G}_{i,n}^a\mathbf{\underline{v}}_{i,n}^a
\end{align}
Likewise, for the prediction error, combining \eqref{as1} and \eqref{augpx_est} gives
\begin{align}\label{augPredError}
	\mathbf{e}_{i,n|n-1}^a &=\mathbf{F}_{n-1}^a\mathbf{e}_{i,n-1|n-1}^a + \mathbf{w}_{n}^a
\end{align}
while the diffused state estimation error can be expressed as
\begin{align}\label{augDiffError}
	\mathbf{e}_{i,n|n}^a 	&= \mathbf{x}_{n}^a - \sum_{k \in \mathcal{N}_i} c_{k,i}\mathbf{\widehat{\underline{x}}}_{k,n|n}^a \nonumber\\
	&= \sum_{k \in \mathcal{N}_i} c_{k,i}\mathbf{\underline{e}}_{k,n|n}^a
\end{align}
Substituting \eqref{augLocalError} and \eqref{augPredError} into \eqref{augDiffError} and using $\mathbf{M}_{k,n|n}^a(\mathbf{M}_{k,n|n-1}^a)^{-1} = \mathbf{I} - \mathbf{G}_{k,n}^a\mathbf{\underline{H}}_{k,n}^a$, we have
\begin{align}\label{augDiffError2}
	\mathbf{e}_{i,n|n}^a 	= &\sum_{k \in \mathcal{N}_i} c_{k,i}\Big[
	\big(\mathbf{I} + \mathbf{G}_{k,n}^a\mathbf{\underline{H}}_{k,n}^a\big)\mathbf{F}_{n-1}^a\mathbf{e}_{k,n-1|n-1}^a \nonumber\\
	&+ \big(\mathbf{I} + \mathbf{G}_{k,n}^a\mathbf{\underline{H}}_{k,n}^a\big)\mathbf{w}_{n}^a
	 + \mathbf{G}_{k,n}^a\mathbf{\underline{v}}_{k,n}^a \Big]	\nonumber\\
	= &\sum_{k \in \mathcal{N}_i} c_{k,i}\Big[
	\mathbf{M}_{k,n|n}^a(\mathbf{M}_{k,n|n-1}^a)^{-1}\mathbf{F}_{n-1}^a\mathbf{e}_{k,n-1|n-1}^a \nonumber\\
	&+ \mathbf{M}_{k,n|n}^a(\mathbf{M}_{k,n|n-1}^a)^{-1}\mathbf{w}_{n}^a
	 + \mathbf{G}_{k,n}^a\mathbf{\underline{v}}_{k,n}^a \Big]	 
\end{align}
Upon taking the statistical expectation, the recursion in \eqref{augDiffError2} leads to a closed form expression for the mean error of the D-ACKF algorithm, given by
\begin{align}\label{MeanaugDiffError2}
	E\{\mathbf{e}_{i,n|n}^a\}	&= \sum_{k \in \mathcal{N}_i} c_{k,i}
	\mathbf{M}_{k,n|n}^a(\mathbf{M}_{k,n|n-1}^a)^{-1}\mathbf{F}_{n-1}^aE\{\mathbf{e}_{k,n-1|n-1}^a\} \nonumber\\
	&= \mathbf{0}
\end{align}
\newline
\textbf{Remark \#6:} Equation \eqref{MeanaugDiffError2} demonstrates that the D-ACKF is an unbiased estimator of general complex processes, exhibiting both proper and improper statistics.,

To derive the mean square error for the D-ACKF, we shall make the following assumptions, commonly used in the analysis of distributed state space estimators.
\newline\newline
\textbf{Assumption \#1:} \emph{Convergence.} All the nodes (local Kalman filters) converge to the same state value by using their neighbourhood data, that is, there are no faulty nodes in the system. 
\newline\newline
\textbf{Assumption \#2:} \emph{Time invariance.} The state space model \eqref{AUGstatespace} is time invariant, that is, $\mathbf{F}_{n} = \mathbf{F}$, $\mathbf{H}_{i,n} = \mathbf{H}$, $\mathbf{Q}_{n} = \mathbf{Q}$ and $\mathbf{R}_{i,n} = \mathbf{R}$, and the state transition matrix $\mathbf{F}$ is stable. This is a standard assumption for the steady-state analysis of Kalman filters.

It then follows that $\lim_{n\rightarrow\infty}\mathbf{M}_{i,n|n}^a = \mathbf{M}_{i,n-1|n_1}^a = \mathbf{M}^a$ for $i \in \{1,\ldots, N\}$, that is, the matrix $\mathbf{M}_{i,n|n}^a $ is also time invariant at steady state.
Next, we define the following terms for convenience of notation:
\begin{align*}
	\mathcal{E}_{i,n|n} &= \begin{bmatrix}\mathbf{e}_{i_1,n|n}^{aT}, \mathbf{e}_{i_2,n|n}^{aT}, \ldots , \mathbf{e}_{i_M,n|n}^{aT} \end{bmatrix}^T	\in \mathbb{C}^{2ML}  \\	
	\mathcal{\underline{W}}_{i,n} &= \begin{bmatrix} \mathbf{w}_{n}^{aT}, \mathbf{w}_{n}^{aT}, \ldots , \mathbf{w}_{n}^{aT} \end{bmatrix}^T	\in \mathbb{C}^{2ML} \\
	\mathcal{\underline{V}}_{i,n} &= \begin{bmatrix} \mathbf{\underline{v}}_{i_1,n}^{aT}, \mathbf{\underline{v}}_{i_2,n}^{aT}, \ldots , \mathbf{\underline{v}}_{i_M,n}^{aT} \end{bmatrix}^T \\	
	\mathcal{\underline{A}}_{i,n} &= \begin{bmatrix} c_{i_1,i}\mathbf{F}_{n-1}^{aT}(\mathbf{M}_{i_1,n|n-1}^a)^{-T} \mathbf{M}_{i_1,n|n}^{aT}\\ c_{i_2,i}\mathbf{F}_{n-1}^{aT}(\mathbf{M}_{i_2,n|n-1}^a)^{-T} \mathbf{M}_{i_2,n|n}^{aT}\\ \vdots \\ c_{i_M,i}\mathbf{F}_{n-1}^{aT}(\mathbf{M}_{i_M,n|n-1}^a)^{-T} \mathbf{M}_{i_M,n|n}^{aT} \end{bmatrix}^T \\	
	\mathcal{\underline{B}}_{i,n} &= \begin{bmatrix} c_{i_1,i}(\mathbf{M}_{i_1,n|n-1}^a)^{-T} \mathbf{M}_{i_1,n|n}^{aT}\\ c_{i_2,i}(\mathbf{M}_{i_2,n|n-1}^a)^{-T} \mathbf{M}_{i_2,n|n}^{aT}\\ \vdots \\ c_{i_M,i}(\mathbf{M}_{i_M,n|n-1}^a)^{-T} \mathbf{M}_{i_M,n|n}^{aT} \end{bmatrix}^T \\
	\mathcal{\underline{G}}_{i,n} &= \begin{bmatrix} c_{i_1,i}\mathbf{G}_{i_1,n}^a, c_{i_2,i}\mathbf{G}_{i_2,n}^a, \ldots , c_{i_M,i}\mathbf{G}_{i_M,n}^a \end{bmatrix}
\end{align*}
where $\{i_1, i_2, \ldots, i_M\} \in \mathcal{N}_i$, and $M = |\mathcal{N}_i|$ is the number of nodes in the neighbourhood $\mathcal{N}_i$. Based on \eqref{augDiffError2}, the mean square error $\mathbf{\Sigma}^a_{i,n} = E\{\mathbf{e}_{i,n|n}^a\mathbf{e}_{i,n|n}^{aH}\}$ at the node $i$ then becomes 
\begin{align}\label{MSE_DACKF}
	\mathbf{\Sigma}^a_{i,n}	\!=\! \mathcal{\underline{A}}_{i,n} \mathcal{M}_{i,n-1}\mathcal{\underline{A}}_{i,n}^H \!+\! \mathcal{\underline{B}}_{i,n}\mathcal{\underline{Q}}_{i,n}\mathcal{\underline{B}}_{i,n}^H \!+\! \mathcal{\underline{G}}_{i,n}\mathcal{\underline{R}}_{i,n}\mathcal{\underline{G}}_{i,n}	
\end{align}
where $\mathcal{M}_{i,n} = E\{\mathcal{E}_{i,n|n}\mathcal{E}_{i,n|n}^H \}$ is the neighbourhood error covariance matrix, $\mathcal{\underline{Q}}_{i,n} = E\{\mathcal{\underline{W}}_{i,n}\mathcal{\underline{W}}_{i,n}^H \}$ and $\mathcal{\underline{R}}_{i,n} = E\{\mathcal{\underline{V}}_{i,n}\mathcal{\underline{V}}_{i,n}^H \}$.
\newline\newline
\textbf{Remark \#7:} Under Assumption \#2, the covariance matrices $\mathcal{\underline{Q}}_{i,n} = \mathcal{\underline{Q}}_{i}$ and $\mathcal{\underline{R}}_{i,n} = \mathcal{\underline{R}}_{i}$ are time invariant, while as $\lim_{n\rightarrow\infty}$ the terms $\mathcal{\underline{A}}_{i,n} = \mathcal{\underline{A}}_{i}$, $\mathcal{\underline{B}}_{i,n} = \mathcal{\underline{B}}_{i}$, $\mathcal{\underline{G}}_{i,n} = \mathcal{\underline{G}}_{i}$ also become time invariant. Then under Assumption \#1, that is, provided all the nodes in the network converge to the same steady state value, the remaining error covariance term $\mathcal{M}_{i,n}$ also converges. 
%

Further, observe that, based on \eqref{augDiffError}, the MSE can alternatively be expressed as
\begin{align}\label{MSE_DACKF2}
	\mathbf{\Sigma}^a_{i,n}	= \sum_{j \in \mathcal{N}_i} \sum_{k \in \mathcal{N}_i} c_{j,i}c_{k,i} \mathbf{\underline{\Gamma}}_{jk,n}^a
\end{align}
where $\mathbf{\underline{\Gamma}}_{jk,n}^a = E\{\mathbf{\underline{e}}_{j,n|n}^a\mathbf{\underline{e}}_{k,n|n}^{aH}\}$ is the cross-correlation matrix between the neighbourhood errors. 
\newline\newline
\textbf{Remark \#8:} Since $\sum_{j \in \mathcal{N}_i} \sum_{k \in \mathcal{N}_i} c_{j,i}c_{k,i} = 1$, the MSE at any node $i$ is then upper bounded by the MSE of the node in its neighbourhood with the worst MSE, that is 
\begin{align}\label{MSE_DACKF3}
	\max\{\mathbf{\Sigma}_{i,n}^a\}	= \max_{k \in \mathcal{N}_i}\{\text{tr}(\mathbf{\underline{\Gamma}}_{kk,n}^a)\}
\end{align}
where $\text{tr}(\cdot)$ is the matrix trace operator. 

It then follows that at any time instant $n$, the upper bound for the average MSE of the whole distributed network is the MSE of the node with the highest MSE in the network.

From Remarks \#6, \#7 and \#8, the D-ACKF converges both in the mean and mean square sense, hence it is a consistent estimator, while its MSE performance is upper bounded by the worst performing node in the network. 
%
%
\section{Application Examples}
To illustrate the advantages of the widely linear D-ACKF over its  strictly linear D-CKF counterpart, the following case studies were conducted: A) filtering of a noisy complex-valued autoregressive process; B) estimating and tracking the position of a projectile in two dimensions.
%
\subsection{Filtering an Autoregressive Process}
\begin{figure}[t]
  \centering  
  \includegraphics[trim = 4mm 5mm 4mm 12mm,scale=0.6]{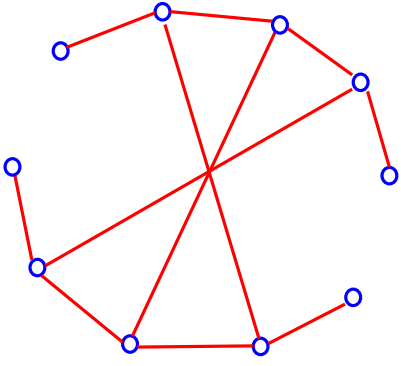}
  \caption{A distributed network with $N=10$ nodes used in the simulations.}
  \label{fig:NetwotkTopologyN10}
\end{figure}
Consider a distributed network consisting of $N=10$ nodes (see Figure \ref{fig:NetwotkTopologyN10}), used for filtering the complex autoregressive (AR) process defined as
\begin{eqnarray*}\label{ar1}
	z_n  =  1.2z_{n-1} - 0.8z_{n-2} + u_n 
\end{eqnarray*}
For rigour, $u_n$ is a noncircular white complex Gaussian driving noise with variance $E\{|u_{n}|^2\} = 2$ and a varying pseudovariance $E\{u^2_{n}\}$. For each node $i$, the observation equation was a noisy measurement of the autoregressive output, that is
\begin{eqnarray*}\label{obs_ar1}
	y_{i,n} = z_{n} + v_{i,n} 
\end{eqnarray*}
where $v_{i,n}$ is the complex Gaussian white observation noise associated with node $i$, while the variances, pseudovariances and cross-correlations of the observation noises were $R_{i,n} = E\{|v_{i,n}|^2\} = 4 + 1/\sqrt{i}$, $U_{i,n} = E\{v^2_{i,n}\}$ and $R_{ik,n} = E\{v_{i,n}v_{k,n}^*\} = 4$ for $i,k \in \{1,2,\ldots,N\}$ and $i\neq k$. 
\newline\newline
\textbf{Remark \#9:}  Observe that the nodes in network experience correlated observation noises with different variances, modelled through the term $1/\sqrt{i}$ in the expression for $R_{i,n}$, where $i = \{1, \cdots, N\}$ is the node index.
\begin{figure}[t]
  \centering  
  \subfloat[Noncircular state noise ]{\label{fig:D-ACKF_noncirc_state}\includegraphics[scale=0.65]{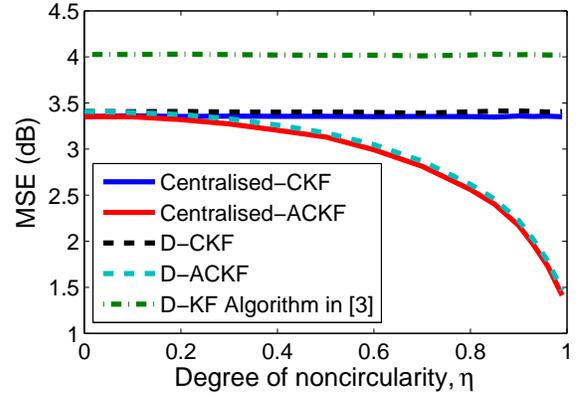}}\\
  \subfloat[Noncircular observation noise]{\label{fig:D-ACKF_noncirc_obs}\includegraphics[scale=0.65]{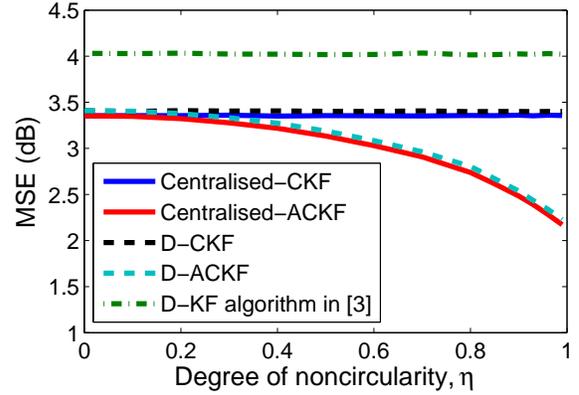}}
  \caption{Steady state performance comparison for filtering the AR($2$) process in the cases of: (a) circular observation noises and a noncircular driving noise with varying degrees of noncircularity; (b) circular state noise and noncircular observation noises with varying degrees of noncircularity, whereby all nodes have same degree of observation noise noncircularity.}
  \label{fig:Circ_D-ACKF}
\end{figure}

In the simulations, we used the ratio of the magnitude of pseudocovariance to covariance, that is $\eta_u =  |E\{u^2\}|/E\{|u|^2\}$, as a measure for the degree of circularity of a (zero-mean) complex signal $u = u_r + ju_i$, where a signal is circular for $\eta_u=0$ and maximally noncircular for $\eta_u=1$. The average mean square errors (MSEs) of all the nodes were used for a quantitative assessment of performance in a nearest neighbour diffusion scheme, which is as follows \cite{Cattivelli_Sayed_IEEETranAC_Dist_KF_2010}. Let $|\mathcal{N}_k|$ denote the number of a neighbours (including itself) of node $k$; to compute the diffused state estimate for node $i$, the weight associated with a neighbour $k$ is proportional to $|\mathcal{N}_k|$, that is
\begin{eqnarray*}
	c_{k,i} = \left\{\begin{array}{l l}
    |\mathcal{N}_k|/\alpha_i & \quad \text{if $k \in \mathcal{N}_i$}\\
    0 & \quad \text{otherwise}\\
  \end{array} \right. 
\end{eqnarray*}
where $\alpha_i = \sum_{k \in \mathcal{N}_i} |\mathcal{N}_k|$ is a normalisation parameter for node $i$ which ensures that $\sum_{k \in \mathcal{N}_i} c_{k,i} = 1$.

Figure \ref{fig:Circ_D-ACKF} compares the steady state performance of the diffusion Kalman filter in \cite{Cattivelli_Sayed_IEEETranAC_Dist_KF_2010} (Algorithm 2), D-CKF and D-ACKF algorithms, along with the centralised versions of the D-CKF and D-ACKF (Centralised-CKF and Centralised-ACKF), with access to the observation data from all the nodes at each time instant. Figure \ref{fig:D-ACKF_noncirc_state} illustrates the results for circular observation noises ($U_{i,n} = E\{v^2_{i,n}\} = 0$ for $i = 1,2,\ldots,N$) and a state (driving) noise with various degrees of noncircularity, whereas the results for a noncircular observation noise with a circular state noise ($P_n = 0$) are shown in Figure \ref{fig:D-ACKF_noncirc_obs}. The variances of the state and observation noises were kept constant throughout, and only their pseudocovariances (degree of circularity) where changed. The results illustrate that for second order circular (proper) state and observation noises ($\eta_w = 0$ and $\eta_{v_i} = 0$), the strictly linear D-CKF and widely linear D-ACKF algorithms have identical performances, conforming with the analysis and Remark \#3, while for noncircular noises ($\eta_w \neq 0$ and $\eta_{v_i} \neq 0$) the D-ACKF offered superior performance, as it catered for the pseudocovariances. Moreover, D-ACKF had decreasing MSE for an increasing degree of noise noncircularity, while D-CKF was unaffected by changes in the noncircularity of the noises, as it is not designed to recognise noncircular signals. 

The performance comparison between the Centralised-CKF and centralised-ACKF algorithms also shows a similar trend, with centralised-ACKF offering better performance for noncircular signals. The D-CKF and D-ACKF algorithms outperformed the diffusion Kalman filter in \cite{Cattivelli_Sayed_IEEETranAC_Dist_KF_2010} (Algorithm 2), because they cater for the cross-correlations between the the observation noises ($R_{ik,n} = E\{v_{i,n}v_{k,n}^*\}$), and only marginally underperformed compared with their centralised counterparts. Observe that for uncorrelated nodal observation noises with circular state and observation noises, the D-CKF, D-ACKF and the diffusion Kalman filter in \cite{Cattivelli_Sayed_IEEETranAC_Dist_KF_2010} will have identical performances.
%
\subsection{Projectile Tracking}
We next considered the problem of estimating and tracking the position of a projectile in two dimensions, where only noisy measurements of its position are available. Let $(x_{n} , y_{n})$ and $(\dot{x}_{n} , \dot{y}_{n})$ denote the position and velocity vectors of the projectile at time instant $n$, respectively, then the corresponding complex valued distributed state space model for the system is given by
\begin{subequations}\label{complexstatespace11}
\begin{eqnarray*}
	\mathbf{x}_n &=& \mathbf{F}\mathbf{x}_{n-1} - j\mathbf{K}g + \mathbf{K}w_n \label{complexStateEqn} \\
	z_{i,n} &=& \mathbf{h}\mathbf{x}_n + v_{i,n} \label{complexObsEqn}
\end{eqnarray*}
\end{subequations}
where:
\begin{itemize}
	\item $\mathbf{x}_n = \begin{bmatrix} x_{n} + jy_{n} & \dot{x}_{n} + j\dot{y}_{n} \end{bmatrix}^T$ is the projectile state vector, and $g=9.8m/s^2$ is the gravitational acceleration;
	\item $\mathbf{F}$, $\mathbf{K}$ and $\mathbf{H}$ are time-invariant matrices and vectors defined as
				\begin{eqnarray*}			
					 \mathbf{F} = \begin{bmatrix} 1 & T \\ 0 & 1  \end{bmatrix} 
					 \text{,} \quad 
					 \mathbf{K} = \begin{bmatrix} \frac{T^2}{2} \\T \end{bmatrix} 
					 \quad \text{and} \quad 
					 \mathbf{h} = \begin{bmatrix} 1 & 0 \end{bmatrix} 					 
				\end{eqnarray*}
				where $T$ is the sampling interval;	
	\item $z_{i,n}$ is the observation at node $i$;
	\item $w_n$ is the zero mean state noise (used to account for modeling inaccuracies), whereas, $\mathbf{v}_{i,n}$ is the zero mean observation noise at node $i$.
\end{itemize}

\begin{figure}[t]
  \centering  
  \includegraphics[trim = 4mm 8mm 4mm 25mm,scale=0.6]{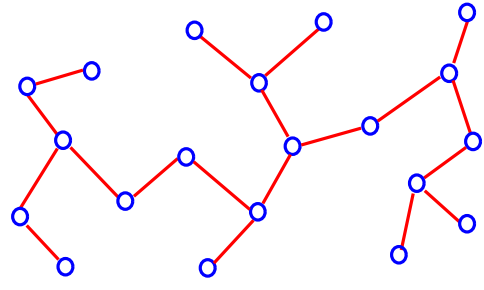}
  \caption{A distributed network with $N=20$ nodes  used in the simulations..}
  \label{fig:NetwotkTopologyN20}
\end{figure}

To illustrate the benefits of the proposed distributed algorithms, we considered a scenario with $N=20$ nodes connected as in Figure \ref{fig:NetwotkTopologyN20}, where a projectile was launched into the air with an initial velocity $(20, 10)$m/s, from location $(0, 0)$m. The sampling interval was set to $T=0.05$s, and the mean square errors (MSEs) of the different algorithms were computed by averaging $1000$ independent trials. The state and observation noises were noncircular random processes, both with a degree of noncircularity of $\eta=0.85$, and their respective distributions were defined as 
\begin{eqnarray*}
	w_{n} \sim  \mathcal{N}(0,5) \qquad \quad
	v_{i,n} \sim  \mathcal{N}(0,1+2\sqrt{i}) 
\end{eqnarray*}
where the observation noise cross-correlations were set to $E\{v_{i,n}v_{k,n}^*\} = 1$ for $i,k \in \{1,2,\ldots,N\}$ and $i\neq k$. 
\begin{figure}[t]
  \centering 
 \includegraphics[scale=0.65]{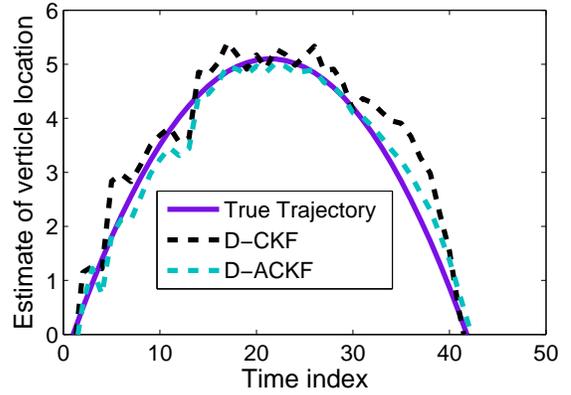}
  \caption{Sample average estimate (of all the nodes) of the vertical position for the diffusion algorithms.}
  \label{fig:D-ACKF_VerticalPosEst}
\end{figure}
\begin{figure}[t]
  \centering  
  \includegraphics[scale=0.65]{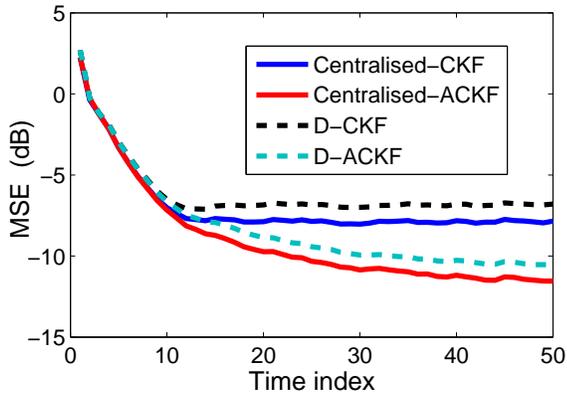}
  \caption{Transient performance of the centralised and diffusion algorithms.}
  \label{fig:MSE_Projectile}
\end{figure}

A sample simulation run for the two diffusion algorithms D-CKF and D-ACKF is shown in Figure \ref{fig:D-ACKF_VerticalPosEst}, while Figure \ref{fig:MSE_Projectile} illustrates the enhanced performance of the widely linear D-ACKF in estimating the projectile location compared with the strictly linear D-CKF. The (strictly linear) centralised-CKF and (widely linear) centralised-ACKF were able to outperform their distributed counterparts D-CKF and D-ACKF, respectively, due to their use of the full network observation data. However, this requires a high communication overhead, compared with that required for the diffusion algorithms, such as the scenario in the sparsely connected network shown in Figure \ref{fig:NetwotkTopologyN20}. 
%
%
\section{Conclusions}
Distributed complex state space estimation has been addressed in the context of collaborative networks for the general case of noncircular states, observations, and state and observation noises. The  distributed (widely linear) augmented complex Kalman filter (D-ACKF) algorithm has been introduced for the sequential state estimation of both second-order circular (rotation invariant) and noncircular (rotation dependent) signal distributions, within a framework which caters for correlated nodal observation noises. We have also analysed its performance, and have shown that it provides unbiased and consistent estimates and enhanced performance for noncircular signals, compared with the distributed complex Kalman filter (D-CKF). Computational complexity issues have been addressed through duality analysis between D-ACKF and a bivariate real valued distributed Kalman filter. Simulations using both circular and noncircular signals illustrate the performance gains of the proposed solutions. 
%
\section*{Acknowledgment}
The authors wish to thank Prof. Ali H. Sayed for his valuable feedback. This work is part of the University Defence Research Centre (UDRC) at Imperial College London, supported by the MoD and DSTL.
%

\bibliographystyle{IEEEtran} 

\bibliography{bibliography}

\begin{thebibliography}{10}
\providecommand{\url}[1]{#1}
\csname url@samestyle\endcsname
\providecommand{\newblock}{\relax}
\providecommand{\bibinfo}[2]{#2}
\providecommand{\BIBentrySTDinterwordspacing}{\spaceskip=0pt\relax}
\providecommand{\BIBentryALTinterwordstretchfactor}{4}
\providecommand{\BIBentryALTinterwordspacing}{\spaceskip=\fontdimen2\font plus
\BIBentryALTinterwordstretchfactor\fontdimen3\font minus
  \fontdimen4\font\relax}
\providecommand{\BIBforeignlanguage}[2]{{%
\expandafter\ifx\csname l@#1\endcsname\relax
\typeout{** WARNING: IEEEtran.bst: No hyphenation pattern has been}%
\typeout{** loaded for the language `#1'. Using the pattern for}%
\typeout{** the default language instead.}%
\else
\language=\csname l@#1\endcsname
\fi
#2}}
\providecommand{\BIBdecl}{\relax}
\BIBdecl

\bibitem{Stadter_Dist_Spacecraft_IEEEMag_2002}
{P. A. Stadter}, {A. A. Chacos}, {R. J. Heins}, {G. T. Moore}, {E. A. Olsen},
  {M. S. Asher}, and {J. O. Bristow}, ``Confluence of navigation,
  communication, and control in distributed spacecraft systems,'' \emph{IEEE
  Aerospace and Electronic Systems Magazine}, vol.~17, no.~5, pp. 26 --32, May
  2002.

\bibitem{Mandic_Fusion_Book_2008}
{D. P. Mandic}, {M. Golz}, {A. Kuh}, {D. Obradovic}, and {T. Tanaka},
  \emph{Signal Processing Techniques for Knowledge Extraction and Information
  Fusion}.\hskip 1em plus 0.5em minus 0.4em\relax Springer, 2008.

\bibitem{Cattivelli_Sayed_IEEETranAC_Dist_KF_2010}
{F. S. Cattivelli} and {A. H. Sayed}, ``Diffusion strategies for distributed
  {K}alman filtering and smoothing,'' \emph{IEEE Transactions on Automatic
  Control}, vol.~55, no.~9, pp. 2069 --2084, Sept. 2010.

\bibitem{Olfati_Saber_Flocking_2006}
R.~Olfati-Saber, ``Flocking for multi-agent dynamic systems{:} algorithms and
  theory,'' \emph{IEEE Transactions on Automatic Control}, vol.~51, no.~3, pp.
  401 -- 420, March 2006.

\bibitem{Zhou_Dist_Architectures_IEEEMag_2003}
{S. Zhou}, {M. Zhao}, {X. Xu}, {J. Wang}, and {Y. Yao}, ``Distributed wireless
  communication system: {A} new architecture for future public wireless
  access,'' \emph{IEEE Communications Magazine}, vol.~41, no.~3, pp. 108 --
  113, March 2003.

\bibitem{Olfati_Saber_Dist_KF}
{R. Olfati-Saber}, ``Distributed {K}alman filtering for sensor networks,'' in
  \emph{46th IEEE Conference on Decision and Control}, Dec. 2007, pp. 5492
  --5498.

\bibitem{Carli_Dist_KF_IEEEComm_2008}
{R. Carli}, {A. Chiuso}, {L. Schenato}, and {S. Zampieri}, ``Distributed
  {K}alman filtering based on consensus strategies,'' \emph{IEEE Journal on
  Selected Areas in Communications}, vol.~26, no.~4, pp. 622 --633, May 2008.

\bibitem{Khan_Dist_KF_IEEETransSP_2008}
{U. A. Khan} and {J. M. F. Moura}, ``Distributing the {K}alman filter for
  large-scale systems,'' \emph{IEEE Transactions on Signal Processing,},
  vol.~56, no.~10, pp. 4919 --4935, Oct. 2008.

\bibitem{Lopes_Sayed_TSP_Dist_LMS_2008}
{C. G. Lopes} and {A. H. Sayed}, ``Diffusion least-mean squares over adaptive
  networks: Formulation and performance analysis,'' \emph{IEEE Transactions on
  Signal Processing}, vol.~56, no.~7, pp. 3122 --3136, July 2008.

\bibitem{Yili_Dist_ACLMS_2011}
{Y. Xia}, {D. P. Mandic}, and {A. H. Sayed}, ``An adaptive diffusion augmented
  {CLMS} algorithm for distributed filtering of noncircular complex signals,''
  \emph{IEEE Signal Processing Letters}, vol.~18, no.~11, pp. 659 --662, Nov.
  2011.

\bibitem{Gao_Dist_Cooperative_IEEETranComm_2011}
{Z. Gao}, H.-Q. Lai, and {K. J. R. Liu}, ``Differential space-time network
  coding for multi-source cooperative communications,'' \emph{IEEE Transactions
  on Communications}, vol.~59, no.~11, pp. 3146 --3157, Nov. 2011.

\bibitem{Mao_Wireless_Comm_IEEETranIFS_2007}
{Y. Mao} and {M. Wu}, ``Tracing malicious relays in cooperative wireless
  communications,'' \emph{IEEE Transactions on Information Forensics and
  Security}, vol.~2, no.~2, pp. 198 --212, June 2007.

\bibitem{Xia_WL_Freq_2011}
{{Y. Xia} and {D. P. Mandic}}, ``Widely linear adaptive frequency estimation of
  unbalanced three-phase power systems,'' \emph{IEEE Transactions on
  Instrumentation and Measurement}, vol.~61, no.~1, pp. 74 --83, Jan. 2012.

\bibitem{Picinbono97}
B.~Picinbono and P.~Bondon, ``{Second-order Statistics of Complex Signals},''
  \emph{IEEE Transactions on Signal Processing}, vol.~45, no.~2, pp. 411--420,
  1997.

\bibitem{Moreno_2008}
{J. Navarro-Moreno}, ``{ARMA} prediction of widely linear systems by using the
  innovations algorithm,'' \emph{IEEE Transactions on Signal Processing},
  vol.~56, no.~7, pp. 3061 --3068, july 2008.

\bibitem{Dini_Class_WLKF_IEEE_TNNLS_2012}
{D. H. Dini} and {D. P. Mandic}, ``A class of widely linear complex {K}alman
  filters,'' \emph{IEEE Transactions on Neural Networks and Learning Systems},
  vol.~23, no.~5, pp. 775 --786, May 2012.

\bibitem{Moreno_2009}
{J. Navarro-Moreno}, {J. Moreno-Kayser}, {R. M. Fernandez-Alcala}, and {J. C.
  Ruiz-Molina}, ``Widely linear estimation algorithms for second-order
  stationary signals,'' \emph{IEEE Transactions on Signal Processing}, vol.~57,
  no.~12, pp. 4930--4935, 2009.

\bibitem{Mandic09}
{D. P. Mandic} and {V. S. L. Goh}, \emph{Complex Valued Nonlinear Adaptive
  Filters: Noncircularity, Widely Linear and Neural Models}.\hskip 1em plus
  0.5em minus 0.4em\relax Wiley, 2009.

\bibitem{Kar_Moura_Dist_KF_IEEETransSP_2011}
{S. Kar} and {J. M. F. Moura}, ``Gossip and distributed {K}alman filtering:
  Weak consensus under weak detectability,'' \emph{IEEE Transactions on Signal
  Processing}, vol.~59, no.~4, pp. 1766 --1784, April 2011.

\bibitem{Hayes96}
M.~H. Hayes, \emph{Statistical Digital Signal Processing and Modeling}.\hskip
  1em plus 0.5em minus 0.4em\relax John Wiley \& Sons, 1996.

\bibitem{Bru_Convergence_1994}
{R. Bru}, {L. Elsner}, and {M. Neumann}, ``Convergence of infinite products of
  matrices and inner-outer iteration schemes,'' \emph{Electronic Transactions
  on Numerical Analysis}, vol.~2, no.~3, pp. 183 --193, Dec. 1994.

\bibitem{Li_Optimal_Dist_fusion_2003}
{X.R. Li}, {Y. Zhu}, {J. Wang}, and {C. Han}, ``Optimal linear estimation
  fusion : Part {I}: Unified fusion rules,'' \emph{IEEE Transactions on
  Information Theory}, vol.~49, no.~9, pp. 2192 -- 2208, Sep. 2003.

\bibitem{Cattivelli_Sayed_ICASSP_Dist_KF_2010}
{F. S. Cattivelli} and {A. H. Sayed}, ``Distributed nonlinear {K}alman
  filtering with applications to wireless localization,'' in \emph{IEEE
  International Conference on Acoustics Speech and Signal Processing (ICASSP)},
  March 2010, pp. 3522 --3525.

\bibitem{Mandic_Duality_ICASSP_2009}
{D. P. Mandic}, {S. Still}, and {S. C. Douglas}, ``Duality between widely
  linear and dual channel adaptive filtering,'' in \emph{IEEE International
  Conference on Acoustics Speech and Signal Processing (ICASSP)}, 2009, pp.
  1745 --1748.

\bibitem{Brandwood_Complex_Grad_1983}
{D.H. Brandwood}, ``A complex gradient operator and its application in adaptive
  array theory,'' \emph{IEE Proceedings on Communications, Radar and Signal
  Processing,}, vol. 130, no.~1, pp. 11 --16, Feb 1983.

\bibitem{Delgado06}
K.~{Kreutz-Delgado}, ``The {C}omplex {G}radient {O}perator and the
  {CR-Calculus},'' Electrical and Computer Engineering, Jacobs School of
  Engineering, University of California, San Diego, Tech. Rep.
  ECE275CG-F05v1.3d, 2006.

\end{thebibliography}
\end{document}